\documentclass[12pt]{article}
\usepackage{amsthm}
\usepackage{amsmath}
\usepackage{lscape}
\usepackage{color}
\usepackage{verbatim}
\usepackage{multirow}
\usepackage{graphicx}
\newtheorem{theorem}{Theorem}

\begin{document}

\def\({{\Bigl(}}
\def\){{\Bigr)}}
\def\[{{\Bigl[}}
\def\]{{\Bigr]}}
\def\H{{\cal H}}
\def\var{{\hbox{var}}}
\def\intt{{\int_0^\infty}}

\vspace{3ex}

\setlength{\baselineskip}{24pt}
\setlength{\parindent}{1em}

 \begin{center}
 {\Large \bf \textsf {Estimation of Dynamic Mixed Double Factors Model in High Dimensional Panel Data}}
 \end{center}
\vspace*{0.1in}

\begin{center}
  Guobin Fang$^\dag$$^{,\ddag}$,   Kani Chen$^*$ and   Bo Zhang$^{\dag,}$\footnote{Address correspondence to  School of Statistics, Renmin University of China. Email: mabzhang@ruc.edu.cn.}
\end{center}

\begin{center} {\it
 $\dag$ Center for Applied Statistics, School of Statistics, Renmin University of China, Beijing  100872, China\\
 $\ddag$ School of Statistics and Applied Mathematics,
 Anhui University of Finance and Economics, Bengbu 233030, Anhui,  China\ \ \\
 $*$ Department of Mathematics, Hong Kong University of Science and Technology, Clear Water Bay, Kooloon, Hong Kong, China}

\end{center}

\vspace{0.1in}

\begin{center}{\bf Abstract}
\end{center}

The purpose of this article is to develop dimension reduction techniques in panel data analysis when the numbers of individuals and indicators { are} large. We use Principal Component Analysis (PCA) method to represent large number of indicators by minority common factors in the factor models. We propose the Dynamic Mixed Double Factor Model (DMDFM for short) to reflect cross section and time series correlation with interactive factor structure. DMDFM { can} not only reduce the dimension of indicators but also { deal with} the time series and cross section mixed effect. Different from other models, mixed factor model have two styles of common factors. The regressors factors reflect common trend and reduce the dimension, { and} error components factors reflect difference and weak correlation of individuals. The results of Monte Carlo simulation show that Generalized Method of Moments (GMM) estimators have good { properties} unbiasedness and consistency. Simulation{ s} also show that the DMDFM can improve prediction power of the models effectively.

\vspace{0.2in}


 \noindent{\it Key words}: Panel data; Dynamic Mixed Double Factor Model; Identification; GMM estimation; Cross-section and time series correlation\\
 \noindent{\it JEL classification}: C23

\newpage

\section{Introduction}

\vspace{0.1in}

\noindent

Processing of large scale data sets of macroeconomic has been one of the cumbersome problems in panel data analysis. Compared with micro panel data, macro panel data { includes} more indicators which { are} usually correlate{ d with} each other. Panel data { covers} cross section and time series data, so the cross correlation results from two aspects: periods and individuals dependency. If these dependenc{ ies} exist, panel data model should be { considered},
regardless of { the source}, { such as in the situation when we compare} economic development across { different} countries or regions. If every country or region { is} regarded as an individual and { observed} by continuous time, cross section and time series correlation { occur since} some of { the} items have { the} same economic structure{ s} and common trends. Similarly, { we may encounter analogical issues} in the micro panel data analysis, { for example the business industry and security market volatility are focused simultaneously when we study the  assets allocation and portfolio management in stock market, which also can be seen as cross section and time series correlation}.
On the other hand,
{ not only correlation between variables and individuals should be considered in high dimensional panel data analysis, how to reduce the number of indicators is also of great importance, which is known as dimension reduction.}

Factor model have been { used} to { analyze} large scale macroeconomic data sets { for a long time}. These macro data sets { consist of} hundreds of indicators and some common trends { can be observed} owing to co-movements of variables, { reflecting the existence of correlation between cross sections}. Chamberlain and Rothschild (1983) { employed} approximate factor structure to study risk free arbitrage portfolio { with weak correlation in} large scale assets { analysis}. They obtain{ ed} the same { conclusions} as Ross (1976) { did in the} arbitrage pricing theory. Forni et al. (2000) propose{ ed} the method of identification and estimation in Generalized Dynamic Factor Model (GDFM). GDFM is the factor model which includes the lag term of factors and cross-correlation of idiosyncratic components.

{ Two parts of the inner structures, i.e. error components and regressors can be considered respectively.} { Some researchers focused on the factor models with factor decomposition of error component only}, see, Ahn, Lee and Schmidt (2001), Moon and Perron (2004), Fan, Fan and Lv (2005), Bai (2009), among many others. They discuss{ ed} unobservable interactive effects of individuals and periods in error components { provided the heterogeneity structure between error and regressors}, which extract error components through factor decomposition. { Besides, factor decomposition with regressors is studied extensively}, see Forni et al. (2000), Stock and Watson (2002), Bai (2003), Anderson and Deistler (2008), etc. In this case, the regressors are expressed { as} two unobservable orthogonal components. Common shocks { are} expressed by minority common factors which { are used to conduct dimension reduction}. Idiosyncratic components { are} expressed by factor loadings to reflect the difference{ s} of individuals. Furthermore, { a complicated case was considered in the} factor decomposition with both of error component and regressors, see, Andrews (2005), Pesaran (2006), among others. They discuss{ ed} the multifactor error structure and cross section { dependency} of individual{} due to the common shock effects.

The lag effects of general dynamic factor model { come from} the lag terms of common factors, i.e. AR or MA processes about common factors. These processes can reflect persistence effect on individual across periods. VAR processes of dynamic factor model also { built} based on lag terms of common factors (eg. Stock and Watson (2005), etc.). Dynamics of common factors derive from regressors' lag effects. Dependent variable in statistical model can be estimated { using} regressors.  Current and past values of regressors will influence dependent variable if we introduce the lag terms of regressors into the model.
{ In the real data analysis}, lag terms of dependent variable { can also} influence current variable values. Stock and Watson (2002) { used} lag effects of dependent variable to forecast macro economy,
but they { didn't consider time series correlation of regressors while transferring} them to lag terms of common factors.

{ Not only in the time series field} does panel data model is used, cross section correlation
{ problems can also be settled}.
{ In this paper}, we propose a Mixed Double Factor Model (MDFM). The double factors refer to factor decomposition with regressors and error components respectively. MDFM { can capture the} structure feature{ s} of panel data with respect to time and individual. We introduce the common factors and factor loadings of regressors and error components to reflect cross section correlation, { and} lag terms of dependent variable { can be seen as} endogenous variables to reflect time series correlation. The Mixed Double Factor Model { including} lag effects of dependent variable is called dynamic Mixed Double Factor Model (DMDFM).

Different from time series and cross section data, panel data include three dimension{ s}: individuals, periods, variables. { In the paper,} We consider { short panel data case first}, where the number of individuals $N$ is larger than periods length $T$. Of course, we will relax this condition at the end of this article. Simultaneously, the number of observable variable $p$ can be larger than $N$ and $T$. Classic statistical modeling methods face multi-collinearity problem. We decompose factors of regressors with principal component analysis (PCA) method. With minority common factors (factor scores) representing large number of explanatory variables, we reduce the number of indicators and parameters to be estimated. On the other hand, common factors reflect correlation among variables.

DMDFM include lag terms of dependent variable in the right hand side (RHS), { and they are correlated} with common factors of regressors and error component. So, we use generalized method of moments (GMM) to estimate the model. Arellano and Bover (1995) { studied} the linear moments conditions and choose the optimal weighting matrix in GMM estimation of dynamic panel data. DMDFM have more complicated structure than classic dynamic pane data model because they include double factors. In this case, The choice of optimal instrumental variables { is} very important. We divide the { processes of DMDFM estimation} into two steps. Firstly, we obtain idiosyncratic component correlated with regressors via GMM estimation, and then PCA method is used to decompose them, { the result of which will be applied} into origin model. Secondly, we { make transformations of} the model { and} estimate the new model with error factors by GMM. By two step iterative method we acquire the uniform optimal estimators. The results of two step estimation can be used to predict the future values of dependent variable.

The rest of this article is organized as follows. Section 2 give some notation{ s} and
the construction processes of DMDFM. Specification and assumptions of DMDFM  will be given in section 3. Section 4 discuss two important problems in DFDFM, one { of which} is { the choice} of factors number, { and} the other { one} is the choice of estimation method. Simulation results will be given in section 5, in which we will simulate the data generation processes of DMDFM. Some conclusion and remarks are provided in section 6.

\vspace{3ex}

\section{Panel data dynamic Mixed Double Factor Model}

\vspace{1ex}
\subsection{Panel data factor model }

In panel data model, let $X_{it}$ and $Y_{it}$ denote the observed value of regressors and responsor on the $t$th period across the $i$th individual, $i$=1,$\cdots$,$N$; $t$=1,$\cdots$,$T$. $X_{it}$ is a $p$ dimensional column vector, $p$ is the number of regressors. Hsiao (2003) consider{ ed} the following model, the slope coefficients { of which} are constant and { intercept} term varies over individuals and time:
$$Y_{it}=\alpha_{it}+\sum_{k=1}^{p}\beta_{k}X_{kit}+u_{it}$$

If the { intercept} terms of above model are regarded as covariances, then the model can be rewritten as matrix form:
$$Y_{it}=X_{it}^{'}B+u_{it} \eqno(1)$$
where $B$ is $p$$\times$1 vector to be estimated, $u_{it}$ is random error term.

Pesaran (2006) propose{ ed a} estimation { method} and { gave the estimators' statistical} inference of linear heterogeneous panel data through multi-factors error structure model:
$$Y_{it}=A_{i}^{'}D_{t}+X_{it}^{'}B+u_{it}$$
where the error term { has a} multi-factors error structure:
$$u_{it}=G_{t}\Gamma_{i}^{'}+\epsilon_{it} \eqno(2)$$
where, $G_{t}$ is unobservable common effects, $\epsilon_{it}$ is { the} individual idiosyncratic error. If $G_{t}$ { is correlated} with $X_{it}$, $X_{it}$ can be expressed as linear combination of $G_{t}$, which { is} named { as} common correlated effect (CCE). Bai (2009) consider{ d a} special case when the number{ s} of individuals $N$ and periods $T$ are very large{ .} Factor loadings and common factors are regarded as unobservable parameter of interactive fixed effects model:
$$Y_{it}=X_{it}^{'}B+G_{t}\Gamma_{i}^{'}+v_{it}$$
{ In the above model}, identification, consistency, limiting distribution of the estimators { were} discussed.

In the case of high dimensional panel data analysis, { in order to reduce} individual data dimension and { reflect} panel data dependent structure feature among individuals, Bai (2003) transfer{ ed} the regressors of model (1) by common factors:
$$X_{it}=F_{t}\Lambda_{i}^{'}+e_{it} \eqno(3)$$
{ where} $\Lambda_{i}$ { represents} factor loadings, $F_{t}$ { is a} common factors vector, { and $e_{it}$ is idiosyncratic error. If }the number of common factors is $r$, then $r$ common factors can be written as: $F_{t}\Lambda_{i}^{'}=\lambda_{i1}F_{1t}+\cdots+\lambda_{ir}F_{rt}$.
Here, $\Lambda_{i}$, $F_{t}$ and $e_{it}$ { are} all unobservable. Model (1) { can be} rewritten as:
$$Y_{it}=F_{t}B^{*}+u_{it}^{*} \eqno(4)$$
where $u_{it}^{*}$ is { an} unobservable idiosyncratic error, { uncorrelated} with $F_{t}$.

\bigskip
\subsection{Panel data dynamic Mixed Double Factor Model}

Since time series and cross section correlation may exist simultaneously{ among} the indicator{ s}, we consider { the situation that} correlation { exists} both in regressors and lag terms of dependent variable when { constructing} panel data factor models. Stock and Watson (2002, 2005) discussed specification and estimation { in} multivariate time series dynamic factor model. They use{ ed} it to extrapolate prediction in multivariate time series case, but do not extend it to panel data model. Meanwhile, idiosyncratic error component $u_{it}$ may exist unobservable interactive effects in { in panel data model}. Considering these factors simultaneously, we propose AR(1) dynamic Mixed Double Factor Model with panel data as follows:
$$Y_{it}=Y_{iw}\beta_{L}+F_{it}\beta_{F}+G_{t}\Gamma_{i}^{'}+\epsilon_{it} \eqno(5)$$
where $Y_{it}$ is { a} dependent variable, represent{ ing} observed value on $t$th period across $i$th individual; $Y_{iw}$ is a column vector composed of the lag terms of $Y_{it}$, $w=t-1,\cdots,t-h$; $\beta_{L}$ and $\beta_{F}$ are $h\times1$ and $r\times1$ { parameter vectors} to be estimated, $F_{it}$ is { an} unobservable $1\times r$ common factors vector. Regressors $X_{it}$ can be decompose{ d as}:
$$X_{it}=F_{it}\Lambda^{'}+e_{it} \eqno(6)$$
{ where} $\Lambda$ is { a} $p\times r$ factor loadings matrix, { and }$r$ { are }common factors decompose{ d} from $p$ regressors ($r<p$), { while that in equation (3) is decomposed from $N$ individuals which are different from each other}. Another group common factors $G_{t}$ and correspondent factor loadings $\Gamma_{i}$ are unobservable $1\times s$ vector{ s}, obtained from regression equation:
$$Y_{it}=Y_{iw}\beta_{L}+F_{it}\beta_{F}+u_{it} \eqno(7)$$
Next, { we} decompose factors from idiosyncratic error $u_{it}$ as equation (2). i.e.
$$u_{it}=Y_{it}-Y_{iw}\beta_{L}-F_{it}\beta_{F}=G_{t}\Gamma_{i}^{'}+\epsilon_{it} \quad (i=1, \cdots, N)$$
where $s$ common factors and corresponding factor loadings can be written as: $G_{t}\Gamma_{i}^{'}=\gamma_{i1}G_{1t}+
\cdots+\gamma_{is}G_{st}$.

{ Using} matrix notation, { we} omit subscript of individuals and periods, {} rewrite equation (7) as { a} simplified style:
$$Y=Y_{L}\beta_{L}+F\beta_{F}+G\Gamma^{'}+\epsilon \eqno(8)$$
where $Y$ and $Y_{L}$ { are} $T\times N$ and $T\times N\times h$ matrix respectively; $F$ is a $T\times N\times r$ matrix with $r$ indicator; $G$ and $\Gamma$ { are} $T\times s$ and $N\times s$ matrix{ s} respectively; $\beta_{L}$ and $\beta_{F}$ { are} $h\times1$ and $r\times1$ coefficient vector{ s}.

From model (8) we consider panel data models with interactive effect in time series and cross section dimension. In this model, lag terms $Y_{L}$ reflect{ s} time series correlation. Without loss of generality, we only consider AR(1) model below. In fact, high order autoregressive model { can be analysed similarly} AR(1). { In this article, we propose a panel data factor modeling strtegy} when the number of indicators $p$ is very large. First group factor $F$ { is used} to reduce the dimension of regressors indicator and multi-collinearity among indicators. Second group factor $G$ reflect{ s} interactive effect{ s} of { the} error component. After twice factorization, idiosyncratic error component $\epsilon$ { can} satisfy model assumption.

Model (8) is { a} generalization of many previous approximate factor model{ s}.
{ Bai (2009) proposed a} interactive fixed effect model, { considering} interactive effect in { the} heterogeneity error term. If we regard the first factor decomposition as identical transformation { of} regressors without considering { the} lag effect, the DMDFM become{ s} the interactive fixed effect model. If we only decompose { the} factor to regressors, DMDFM become{ s} classic factor model.

Compared with Pesaran (2006){ 's} multi-error structure model, DMDFM { can handle} both individual effect of regressors and lag effect for dependent variable. In the processes of factor decomposition, if we decompose common factor $F$ and $G$ with { the} same method, DMDFM become{ s the} multi error structure model.

Anderews (2005) propose{ d} the common shocks of cross section regression { which} generalized classic common factor model, but { using that model, the paper only discussed} common shocks to cross section without giving specific form of common factors. If we regard factor decomposition of DMDFM as common shocks, the same conclusion as Andrews should be obtained.

The forecasting idea of DMDFM is slightly different from Stock and Watson (2002) { because} we introduce double style factors to reflect time and individual correlation. DMDFM generalized style of Stock and Watson from multivariate time series to panel data, { and} the more complex factors will be considered.

\bigskip

\section{Identification and assumption of DMDFM}

Generally, we assume that the number{ s} of individual $N$ and periods length $T$ are very large when we investigate high dimensional panel data. We pay more attention to large $N$ and $p$, where the dimension{ s} of individuals and indicators { are} very large. The relative size{ s} of $N$ and $p$ aren't restricted strictly.

The problem of parameters estimation and variable identification derived from not enough restriction condition, in this case the values are not { unique}. { For} factor model, the problem{ s} of proper identification and estimation { is that there exist} more assumptions { compared with} classic panel data model. We apply some assumption condition{ s} to factors and factor loadings, { and} the constraint condition{ s are also applied} to error term, regressors and model (2) (5) (6) (7).

Assumption A: (Identification)

a1. $\Lambda^{'}\Lambda/p\rightarrow I_{r}$.

a2. $E(FF^{'})=\Sigma_{FF^{'}}$, where $\Sigma_{FF^{'}}$ is { a} order $r$ positive diagonal matrix; the subscript of $F_{it}$ is omitted for simplicity.

a3. $\Gamma^{'}\Gamma/N\rightarrow I_{s}$.

a4. $E(G_{t}G_{t}^{'})=\Sigma_{GG^{'}}$, where $\Sigma_{GG^{'}}$ is { a} order $s$ positive definite diagonal matrix.

{ We know that} $F_{it}\Lambda^{'}=F_{it}RR^{-1}\Lambda^{'}$ and $G_{t}\Gamma_{i}^{'}=G_{t}QQ^{-1}\Gamma_{i}^{'}$, where $R$ and $Q$ { are} arbitrary invertible matrix{ es with} order $r$ and $s$. If we do not { add} some constraint conditions to them, decomposition factor of regressors and error terms { won't} be { unique}. Assumption a1 and a2 can { cause} $r^{2}$ restriction{ s} for first group common factors $F_{it}$ and factor loadings $\Lambda$. Assumption a3 and a4 can { lead to} $s^{2}$ restriction{ s} for second group common factors $G_{t}$ and factor loadings $\Gamma_{i}$. Stock and Watson (2002) argue{ d} that  assumption a2 and a4 can ensure covariance stationary if we introduce{ d} lag terms of common factors $F_{it}$ and $G_{t}$ { into} dynamic factor model (5). Bai (2009) propose{ d} some invertible assumption{ s} { in} coefficient matrix { for} identification and estimation { of} parameter $\beta_{L}$ and $\beta_{F}$.

Assumption B: (Factors and factor loadings)

b1. $\|\lambda_{i}\|<\lambda_{max}<\infty$.

b2. $E\|F\|^{4}<\infty$, and $p^{-1}\sum_{p}FF^{'}\stackrel{p}{\rightarrow}\Sigma_{FF^{'}}$, the subscript of $F_{it}$ is omitted for simplicity.

b3. $\|\gamma_{i}\|<\gamma_{max}<\infty$, $E\|G_{t}\|^{4}<\infty$.

Frobenius norm of matrix $F$ is defined as $\|F\|=[tr(F^{'}F)]^{1/2}$, where $tr(F)$ is the trace of matrix. Assumption b1-b3 can assure common factors $F_{it}$ and $G_{t}$ with correspondence factor loadings are not infinity. Bai and Ng (2002) argue{ d} that { the} above factors and factor loadings can ensure factor model { standardization} and improve the efficiency of factor decomposition { in} primitive variable.

Assumption C: (Errors component)

c1. $E(\epsilon_{it})=0$, $Var(\epsilon_{it})=\sigma_{\epsilon}^{2}$, $E(Y_{it}Y_{it+h})=\rho_{i}(h)$,

\qquad$lim_{N\rightarrow\infty}sup_{t}\sum_{N}\|\rho_{i}(h)\| \leq M<\infty$.

c2. $E(Y_{it}Y_{jt})=\tau_{t}(k)$.

c3.  For every(t,s), $E(N^{-1}\sum_{i}|\epsilon_{is}\epsilon_{it}-E(\epsilon_{is}\epsilon_{it})|^{4})\leq M<\infty$

c4. $lim_{N\rightarrow\infty}sup_{i}\sum_{i,j}\sum_{s,t,u,v}\|cov(\epsilon_{is}\epsilon_{it},\epsilon_{ju}\epsilon_{jv})\|\leq M<\infty$

Assumption{ s} of error term and its moments come from three parts: mean, variance, { and} moments condition, { which} are also called weak correlation assumption{ s}. Assumption c1 restrict{ s} weak correlation of time series and mean of error term ruled out by twice factor decomposition, where the weak correlation is ready to the follow discussion of dynamic factor model. Assumption c2 represent{ s} cross section correlation. Assumption c3 give{ s} high order moments condition with uniform bound. Assumption c4 is { the} covariance bound of TS/CS, which is more stricter than c1-c3.

The idiosyncratic error $e_{it}$ and $\epsilon_{it}$ from regressors $X_{it}$ and error term $u_{it}$  must satisfy { the} assumption of factor decomposition, i.e., idiosyncratic errors are mutually independent, mean 0, { and} diagonal covariance matrix with off-diagonal elements 0.

Assumption D: (Dependent variable, common factors and model parameters)

d1. $E(Y_{iw}^{'}G_{t})=\xi, E(G_{t}^{'}F_{it})=\psi$.

d2. $E[G_{t}\epsilon_{it}(G_{t}\epsilon_{it})^{'}]=T^{-1}\sum_{s}\sum_{t}(G_{t}G_{t}^{'}\epsilon_{is}\epsilon_{it})$ (iff $t$ $\rightarrow\infty)$,

\qquad$E[G_{t}F_{it}(G_{t}F_{it})^{'}]=\Sigma_{FG}$, $E[(Y_{iw}^{'}F_{it})^{'}Y_{iw}^{'}F_{it}]=\Sigma_{YF}$, where $\Sigma_{FG}$ and $\Sigma_{YF}$ are block diagonal positive matrix.

d3. $\|\beta_{L}\|<\infty, \|\beta_{F}\|<\infty$.

Assumption D impose on the relationship between regressors and error term, { including the} key condition{ s to be used in parametric estimation}. Assumption d1 reflect{ s the} correlation of regressors in model (5), { while} assumption C { has} given some weak correlation in the other variables. Assumption d2 is very strong which ensure model (5) can be estimated. Assumption d3 restrict{ s} the bound of $\beta_{L}$ and $\beta_{F}$.

Assumption A-D describe inner structure of model (2)-(7), and guarantee that each model can be estimated. We will study how to estimate the model and discuss the asymptotic property of { the} estimator { with} large $N$ and large $p$.

\section{Model estimation}

\subsection{Factor decomposition and { choice of} the number of factors}

DMDFM { does the factor decomposition} twice, so that the method of factor decomposition and the choice of factor number are very important. Many literatures have discussed the choice of lag orders and the number of factors, but the schemes they proposed are only adaptive to lag of factor, e.g., followed by Forni et al.(2000){ 's} generalized dynamic factor model (GDFM), Hallin and Liska (2007) propose{ d} valid information to choose the number of common factors, whose method is based on spectral density matrix decomposition theory. Harding and Nair (2009) exploit{ ed} random matrix theory and Stieltjes transformation { in uniform estimation deriving processes} to determine lag orders and the number of common factors for common shocks component. This is named { as} dynamic scree plot method, where the GDFM is conveyed as follows:
$$R_{t}=\sum_{i=0}^{q}\Lambda_{i}F_{t-i}+\epsilon_{t}$$
where $R_{t}$ is { a} $N\times1$ vector, { and} the dynamic refer{ s} to lag effect of factors, which is different from the dynamic model of dependent variable in this article.

We decompose factors twice in this paper{ . Firstly,} we { use} equation (2) to handle weak correlation and reduce dimension of individual, where common factor { is} composed of common shocks by different individual{ s}. Secondly, we use classic PCA method to decompose factor in equation (2) and (6). We will apply two different methods to choose the number of factors . {  We can use nonparametric scree plot method to choose} the factor number of regressors in model (6) because the common factors of model (6) extracted from large indicators as multivariate analysis, in which factors number determined by scree plot method through contribution rate of variance can reflect indicator information maximize.

\bigskip
\texttt{Remark 1:} We decompose factors on { each period}, and obtain different factor numbers { varying} with periods. It$'$s very important to choose a { unified} number of factors, which can improve analysis efficiency. Here, we choose the maximum contribution rate of variance to determine the number of common factors.

\bigskip
{ Determining} the factors{ 's} number of idiosyncratic error $u_{it}$ is more complicated because they are additional information{ s} after several times transformation. Bai and Ng (2002) propose{ d} { two choice strategies} of number factors for panel data, { and they are} all derived from Mallows (1973) information criterion ($C_{p}$).

One of them is panel data $C_{p}$ criteria ($PC_{p}$), { including} three styles, { among which the basic one is:}
$$PC_{p1}(k)=V(k,\hat{F}_{k})+k\hat{\sigma}^{2}(\frac{N+T}{NT})ln(\frac{NT}{N+T})$$
where $V(k,\hat{F}^{k})=N^{-1}\sum_{i=1}^{N}\hat{\sigma}_{i}^{2}$, and $\hat{\sigma}_{i}^{2}=\hat{\epsilon}_{i}^{'}\hat{\epsilon}_{i}/T$. $PC_{p}$ { is a} minimizing criteria with square sum of error{ s} plus a penalty function. $PC_{p2}$ and $PC_{p3}$ are similar with $PC_{p1}$.

The other one is panel information criteria ($IC_{p}$), { corresponding to} $PC_{p}$. { They} also have three styles, one of { which} is:
$$IC_{p1}(k)=V(k,\hat{F}_{k})+k(\frac{N+T}{NT})ln(\frac{NT}{N+T})$$
The advantage of this criteria is that it doesn't depend on square error $\hat{\sigma}_{i}^{2}$, { which} may extend the application scope. Bai and Ng (2002) { argued} that both $IC_{p}$ criteria and $PC_{p}$ criteria can choose the number of factors in panel data analysis.

$PC_{p}$ and $IC_{p}$ information criteria { can both} be used to factors number choice for panel data. DMDFM decompose factors { for} twice. Equation (6) is a multivariate PCA decomposition, however equation (2) is a panel data factors decomposition of error component{ s}. In the processes of idiosyncratic error $U_{it}$ decomposition, { we use} $PC_{p}$ and $IC_{p}$ minimization criteria { in the choice of factors numbers}. The regressors' factors number will be chosen by variance contribution method or scree plot method.

\subsection{Estimation processes of DMDFM}

The estimation { process} of DMDFM (2)-(7) can be divided into { the} following four steps: Firstly, decompose factors with regressors $X_{it}$; Secondly, estimate model (7); Thirdly, decompose factors with error term $u_{it}$; At last, estimate model (5). The two step estimation and two step factors decomposition are different { from} their realized processes respectively.

At first, we reduce the dimension of multiple indicators of regressors $X_{it}$ from $p$ to $r$ ($r<p$), where the number of factors $r$ is determined by the rate of variance contribution. The results can be expressed as:
$$X_{it}=\tilde{F}_{it}\tilde{\Lambda}^{'}+e_{it} \eqno(9)$$

\bigskip
\texttt{Remark 2:} Common factors $\tilde{F}_{it}$ and factor loadings $\tilde{\Lambda}$ are unobservable, and the information of regressors $X_{it}$ are reflected { by} common factors $\tilde{F}_{it}$. Here, we use factor scores in equation estimation rather than common factors. Factor scores can be obtained by weighting least square or other methods.

\bigskip

Next, { we} substitute $\tilde{F}_{it}$ and $Y_{it}$ $'$s lag terms $Y_{iw}$ into model (7), { and} use Generalized Method of Moments (GMM) to obtain models$'$ initial parameter estimators $\hat{\beta}_{L}$ and $\hat{\beta}_{F}$. Furthermore, { we} calculate the error of model (7) from the results of GMM estimation:
$$\hat{u}_{it}=Y_{it}-\hat{Y}_{it}=Y_{it}-Y_{iw}\hat{\beta}_{L}-F_{it}\hat{\beta}_{F}$$

Then, { we need to} decompose factor with $u_{it}$, { using} $PC_{p}$ and $IC_{p}$ criteria to determine the number of common factors $s$. The results of decomposition can be { expressed} as:
$$\tilde{u}_{it}=\tilde{G}_{t}\tilde{\Gamma}_{i}^{'}+\epsilon_{it} \eqno(10)$$

Finally, { we} substitute the results of twice factor decomposition into model (5), estimate model (5), and obtain the estimation parameters $\tilde{\beta}_{L}$ and $\tilde{\beta}_{F}$ { as well as} the prediction equation:
$$\tilde{Y}_{it}=Y_{iw}\tilde{\beta}_{L}+\tilde{F}_{it}\tilde{\beta}_{F}+\tilde{G}_{t}\tilde{\Gamma}_{i}^{'} \eqno(11)$$

When { estimating} model (5), we can get $\tilde{\Gamma}^{'}\tilde{\Gamma}/N=I_{r}$ through assumption a3 and a4, which provide the identification condition of common factors and factor loadings. At the same time, equation (10) provide the result of decomposition for common factors $G_{t}$ and factor loadings $\Gamma_{i}$, so $\tilde{G}_{t}\tilde{\Gamma}_{i}^{'}$ in equation (11) can be observable. We consider the correlation { between} lag terms and regressors when we estimate model (5). Thus, we { employ} GMM to estimate the parameters of model (11).

{ The above} four step estimation method{ s} include two step factor decomposition and two step model estimation. { The} first step factor decomposition { make the goal of indicators' dimension reduction realized}, { identifying the} typical factors and their scores { to} represent all covariates and their values. { The} second step factor decomposition mainly reflect idiosyncratic and interactive effects of individuals and periods. { In the following, we consider the} two step estimation procedure{ s} provided in the model. { The} first step extract idiosyncratic errors to decompose factor of interactive effect{ s}. { The} second step { gives consistent} estimator of model (5). { The} choice of correct estimation methods of given model is very important, { otherwise} we will get { an} incorrect estimation result. Here, we consider applying generalized moments method (GMM).

\subsection{Realization of estimation processes}

Model (5) include lag term of common factors and dependent variable, therefore { it's difficult to use} maximum likelihood estimation method to get strong uniform convergence results. Arellano and Bond (1991) consider{ d} GMM estimation { in} individual random effect panel data autoregressive model with independent strict exogenous variables and predetermined variables. Arellano and Bover (1995) develop{ d} the method of instrumental variable selection through GMM estimation in panel data model which include predetermined variable, { and} they characterize the valid transformations for exogenous variables. GMM is more flexible for the panel data model estimates with lags and exogenous variables, and it { can also} be regarded as a { consistent} parameter estimation method for DMDFM.

We need { to} determine moment condition{ s} and choose optimal instrumental variable if we { use} GMM to estimate panel data DMDFM. Without loss of generality, we only discuss the AR(1) process of dependent variable below. Here, model (5) can be written as:
$$Y_{it}=Y_{it-1}\rho+F_{it}\beta_{F}+G_{t}\Gamma_{i}^{'}+\epsilon_{it}\eqno(12)$$
Because { the} common factor $G_{t}$ and factor loading $\Gamma_{i}$ { are} obtained from decomposition of equation (10), $G_{t}$ and $\Gamma_{i}$ are observable when { estimating} model (12), { the estimators of which are} denoted by $\tilde{G_{t}}$ and $\tilde{\Gamma_{i}}$, and model (12) { becomes:}
$$Y_{it}=Y_{it-1}\rho+F_{it}\beta_{F}+\tilde{G_{t}}\tilde{\Gamma}_{i}^{'}+\epsilon_{it}\eqno(13)$$
For simplicity, we still use notation $\epsilon_{it}$ { representing} error component{ s} in model (13). { Following} the inspiration of Arellano and Bond (1991), Hsiao (2003), instrumental variables maybe choose lag terms of dependent variable (predetermined variable) and exogenous variables. For the model (13), the choice of instrumental variables should { be correlated} with explanatory variable and { orthogonal} with the residual terms. So, { implementing} first order difference transformation { in} the model (13), we obtain
$$Y_{it}-Y_{it-1}=(Y_{it-1}-Y_{it-2})\rho+(F_{it}-F_{it-1})\beta_{F}+
(\tilde{G_{t}}-\tilde{G_{t-1}})\tilde{\Gamma}_{i}^{'}+\epsilon_{it}-\epsilon_{it-1}$$
Here, $(\tilde{G}_{t}-\tilde{G}_{t-1})\tilde{\Gamma}_{i}^{'}$ is observable scalar variable, { and} it can be combined with constant term when we estimate model (12), { or the model including a constant term in the model (12)}.

\bigskip
\texttt{Remark 3:}We assume { that the} factor decomposition of error component can be substituted into constant terms, so they can be regarded as { a} constant factor amongst common factors $F_{it}$. If { so}, we should replace $F_{it}$ with new notation{ s}. For the sake of brevity, we still use the same notation{ s} as before, but the factorization results of error components are { included} in the error terms of model (13).

\bigskip

The first order difference transformation of model (13) can be written as
$$Y_{it}-Y_{it-1}=(Y_{it-1}-Y_{it-2})\rho+(F_{it}-F_{it-1})\beta_{F}+\epsilon_{it}-\epsilon_{it-1}$$
rewritten as difference operator $\Delta$
$$\Delta Y_{it}=\Delta Y_{it-1}\rho+\Delta F_{it}\beta_{F}+\Delta\epsilon_{it} \eqno(14)$$
{ The lag terms of $Y_{it}$, $Y_{it-2-j}$ ($j=0,1,2,\cdots,t-2$) is} subject to $E[Y_{it-j-2}(Y_{it-1}-Y_{it-2})]\neq0$ and $E[Y_{it-j-2}(\epsilon_{it}-\epsilon_{it-1})]=0$. For the $i$th individual which includes $T(T-1)/2$ moment conditions, { the} difference of the error term, $(\epsilon_{it}-\epsilon_{it-1})$, $t=2$, $\cdots$, $T$, { is denoted as} $\Delta\epsilon_{i}${ . Here} $r$ explanatory variables $F_{it}$ have similar features with $Y_{it-2-j}$,
$$E[F_{it}\Delta\epsilon_{i}]=0, t=1, \cdots, T$$
Thus, we obtain $r\times T \times (T-1)$ moment conditions for $i$th individual, { and} predetermined variables and exogenous variables can determine $T(T-1)/2+r\times T\times(T-1)$ moment equations of residual term. Denotes
$$H_{it}=(Y_{i0},\cdots,Y_{it-2},F_{i1}^{'},\cdots,F_{iT}^{'})^{'}$$
the $T(T-1)/2+r\times T\times (T-1)$ moment equations can be written as:
$$E[H_{it}\Delta\epsilon_{it}]=0, t=2,\cdots,T$$
{ These} moment equations provide some moment conditions to error terms.
For simplicity, { we} omit the subscript $t$ for all variables, and obtain matrix form { of} the model:
$$\Delta Y_{i}=\Delta Y_{i-1}\rho+\Delta F_{i}\beta_{F}+\Delta\epsilon_{i}, i=1, \cdots,N \eqno(15)$$
{ Denote}
$$
Z_{i}=\begin{bmatrix}
  H_{i2} & 0 & \cdots & 0 \\
  0 & H_{i3} & \cdots & 0 \\
  \vdots & \vdots & \ddots & \vdots \\
  0 & 0 & \cdots & H_{iT}
\end{bmatrix}
$$
for the $i$th individual, the previous moment equations can be written as:
$$E[Z_{i}\Delta \epsilon_{i}]=0, i=1, \cdots,N \eqno(16)$$
Because the number of moment equations in equation (16) is $T(T-1)/2+r\times T\times(T-1)$ which is much larger than the number of parameter{ s} to be estimated in model (15), $r+1${ . We} impose some restriction conditions on it. The residual sum of squares of model (15) { is} define as follows:
$$V(\Delta Y,\Delta F;\rho,\beta)=\sum_{i=1}^{N}(\Delta Y_{i}-\Delta Y_{i,-1}\rho-\Delta F_{i}\beta_{F})^{'}(\Delta Y_{i}-\Delta Y_{i,-1}\rho-\Delta F_{i}\beta_{F}) \eqno(17)$$
We can obtain uniform optimal estimator of unknown parameter{ s} through minimizing objective function (17). Too many moment conditions { causes} the moment equations (16) { insoluble}. To acquire valid conditions of { the} parameter estimation, we seek { a} positive definite matrix $A$, { with} which { the} transform objective function (17) { is written as:}
$$\tilde{V}(\Delta Y,\Delta F;\rho,\beta)=\sum_{i=1}^{N}(\Delta Y_{i}-\Delta Y_{i,-1}\rho-\Delta F_{i}\beta_{F})^{'}A(\Delta Y_{i}-\Delta Y_{i,-1}\rho-\Delta F_{i}\beta_{F}) \eqno(18)$$
Through minimizing objective function (18), we can obtain estimators $\hat{\rho}$ and $\hat{\beta}_{F}$ of parameter $\rho$ and $\beta_{F}$, by choosing appropriate positive definite matrix{ s}. The covariance matrix of $Z_{i}\Delta\epsilon_{i}$ is:
$$V_{N}=N^{-1}\sum_{i=1}^{N}E(Z_{i}\Delta\epsilon_{i}\Delta\epsilon_{i}^{'}Z_{i}^{'})$$
{ whose} estimation results can be written as:
$$\hat{V}_{N}=N^{-1}\sum_{i=1}^{N}E(Z_{i}\Delta\hat{\epsilon}_{i}\Delta\hat{\epsilon}_{i}^{'}Z_{i}^{'})$$
{ From} the results of Hansen (1982), { we see that} optimal alternative $A_{O}$ of positive definite matrix $A$ is $\hat{V}_{N}^{-1}$. From previous assumption C, { we know} error term $\epsilon_{it}$ is i.i.d., mean 0, variance $\sigma_{\epsilon}^{2}$, so we have:
$$A_{O}=(N^{-1}\sum_{i=1}^{N}Z_{i}UZ_{i}^{'})^{-1}$$
According to the one step estimation method of Arellano and Bond (1991), known transformation matrix can't extract the information of error term thoroughly. We consider using two-step estimation method, { and} the residual $\hat{\epsilon}_{i}^{(1)}$ of first step estimation to construct transformation matrix $U_{i}=\sum_{i=1}^{N}\hat{\epsilon}_{i}^{(1)}\hat{\epsilon}_{i}^{(1)'}$. Then we minimize objective function (18) and obtain the estimators of $\rho$ and $\beta_{F}$ similar { to} Arellano and Bond (1991):
$$
(\hat{\rho},\hat{\beta}_{F})=((\Delta Y_{-1},\Delta F)^{'}Z^{'}A_{O}Z(\Delta Y_{-1},\Delta F))^{-1}(\Delta Y_{-1},\Delta F)^{'}Z^{'}A_{O}Z\Delta Y \eqno(19)
$$
where $\Delta Y_{-1}$ and $\Delta F$ are $N(T-1)$ vector and $N(T-1)\times r$ matrix respectively, which represent predetermined variables and exogenous variables. These two styles of variables can be estimated respectively
or simultaneously as explanatory variables. The meaning of $A_{O}$ and $Z$ as { mentioned before}, represent optimal choice of transformation matrix and weighted matrix respectively. $Z$ is a block diagonal matrix composed by the instrumental variables.

\subsection{Theory results}

GMM estimation solve population moment equations through sample moment conditions, with regard to the case of over identification{ . We} transform them { for} identification by weighted matrix or transformation matrix $A$. If the optimal weighted matrix and the instrumental variable matrix have been { correctly} chosen, the GMM estimators { have} consistency and asymptotic normality. The sample estimators of { parameters in} equation (19) obtained { by} minimizing objective function (18) can be written as:
$$
(\hat{\rho},\hat{\beta}_{F})=\bigg{\{\[}\sum_{i}(\Delta Y_{i,-1},\Delta F_{i})^{'}Z_{i}^{'}\Big{]}\Big{[}\sum_{i}Z_{i}A_{O}Z_{i}^{'}\Big{]}^{-1}\Big{]}\sum_{i}Z_{i}(\Delta Y_{i,-1},\Delta F_{i})\Big{]}\bigg{\}}^{-1}$$
$$\times\Big{[}\sum_{i}((\Delta Y_{i,-1},\Delta F_{i})^{'}Z_{i}^{'}\Big{]}\Big{[}\sum_{i}Z_{i}A_{O}Z_{i}^{'}\Big{]}^{-1}\Big{]}\sum_{i}Z_{i}\Delta Y_{i},\Big{]}\eqno(20)
$$
{ The} RHS of model (5) include the high order lag terms of dependent variable $Y_{it}$, which be seem as IV in GMM estimation to obtain consistent efficiency estimators of regression parameter. { After the} previous assumption conditions are satisfied, we could draw more general conclusion as below.

\begin{theorem}(Consistency)
Under assumption conditions { A-D}, GMM estimators $\tilde{\beta}_{L}$ and $\tilde{\beta}_{F}$ are the { estimators} of lag terms parameter $\beta_{L}$ and common factor parameter $\beta_{F}$ respectively. Suppose the number of explanatory variables $p$ and period length $T$ are given, when $N\rightarrow\infty$, { the} following conclusions are found:\\
(1)$\tilde{\beta}_{L}-\beta_{L}\rightarrow0,\quad \tilde{\beta}_{F}-\beta_{F}\rightarrow0$\\
(2)$Y_{iw}\tilde{\beta}_{L}+\tilde{F}_{it}\tilde{\beta}_{F}+\tilde{G}_{t}\tilde{\Gamma}_{i}^{'}-
(Y_{iw}\beta_{L}+F_{it}\beta_{F}+G_{t}\Gamma_{i}^{'})\rightarrow0$
\end{theorem}

The proofs of theorem 1 can be found in Appendix A.

The conclusion (1) of theorem 1 indicates that the coefficient estimators of predetermined variables and exogenous explanatory variables converge w.p.1. to  real parameter as sample size tends to $\infty$. Conclusion (2) demonstrates { the} consistency { of the estimators}, { and} more detail { can be seen in the proof}, where the expression of estimators can be { analogized} from equation (19) and (20). The estimation results of model parameter have consistency, so { they can be applied to} extrapolation and prediction.

Suppose random error term $\epsilon_{it}$ is i.i.d., and mean 0, variance $\sigma_{\epsilon}^{2}$, following normal distribution, where optimal transformation matrix $A_{O}$ and weighted matrix $Z$ are chosen in GMM estimation, we obtain { the asymptotic variance of estimator:}
$$
avar(\hat{\rho},\hat{\beta}_{F})=\sigma_{\epsilon}^{2}\bigg{\{}\Big{[}\sum_{i}(\Delta Y_{i,-1},\Delta  F_{i})^{'}Z_{i}^{'}\Big{]}\Big{[}\sum_{i}Z_{i}A_{O}Z_{i}^{'}\Big{]}^{-1}\Big{[}\sum_{i}Z_{i}(\delta Y_{i,-1},\Delta  F_{i})\Big{]}\bigg{\}}^{-1}
\eqno(21)$$

{ We rewrite} objective function (18):
$$
O_{N}=N^{-1}\sum_{i=1}^{N}(\Delta Y_{i}-\Delta Y_{i,-1}\beta_{L}-\Delta F_{i}\beta_{F})^{'}A(\Delta Y_{i}-\Delta Y_{i,-1}\beta_{L}-\Delta F_{i}\beta_{F}) \eqno(22)
$$
{ Calculating the} first order partial derivative to objective function $O_{N}$ with respect to parameter $\beta_{L}$ and $\beta_{L}$, { we have}:
$$R_{L}=\partial O_{N}/\partial \beta_{L} \quad and \quad R_{F}=\partial O_{N}/\partial \beta_{F}$$
where $R(\beta _{L} ,\beta _{F} ) = (\beta _{L}^{'} ,\beta _{F}^{' })'$ are first order partial derivative{ s} with respect to { the parameters} to be estimated, { since} we obtain { the estimators form (20)} via { minimizing} objective function (18), which converge to (19) consistently. Furthermore,{ notice that} random matrix $R$ converge to matrix $R_{1}$ w.p.1., and denote
$$
\Sigma_{1}=(R_{1}^{'}A_{O}R_{1})^{-1}R_{1}^{'}A_{O}D_{1}A_{O}R_{1}(R_{1}^{'}A_{O}R_{1})^{-1}
$$
where $D_{1}$ is { the} asymptotic variance of $\sqrt{N}O_{N}$ when $N\rightarrow\infty$,
$$\sqrt{N}O_{N} \xrightarrow{d} N(0,D_{1}) \eqno(23)$$
{ Here we }assume $\sqrt{N}O_{N}$ converge{ s} in distribution to normal distribution with mean 0.

{ The above analysis are all based} on short panel data ($T<N$){ . Furthermore, we} consider long panel data { whose} periods length $T$ and individual number $N$ tend to infinity simultaneously, and $R \xrightarrow[a.s.]{p} R_{2}${ . Let}
$$
O_{N}=(NT)^{-1}\sum_{i=1}^{N}(\Delta Y_{i}-\delta Y_{i,-1}\beta_{L}-\Delta F_{i}\beta_{F})^{'}A(\Delta Y_{i}-\delta Y_{i,-1}\beta_{L}-\Delta F_{i}\beta_{F})
$$
{ and the other notations remain unchanged. Denote}
$$
\Sigma_{2}=(R_{2}^{'}A_{O}R_{2})^{-1}R_{2}^{'}A_{O}D_{2}A_{O}R_{2}(R_{2}^{'}A_{O}R_{2})^{-1}
$$
when $N,T \rightarrow \infty$, { we} assume
$$\sqrt{NT}O_{N} \xrightarrow{d} N(0,D_{2}) \eqno(24)$$

Under the given correlation assumption{ s}, when periods length $T \rightarrow \infty$, GMM estimators of dynamic double factors model { have} asymptotic normality. The conclusions { can be seen in Theorem 2}.

\begin{theorem}(CLT)
Given some positive matrix{ es} $\Sigma_1$ - $\Sigma_{2}$, under assumption conditions, the conclusion{ s are} as follows:\\
(1) Explanatory variables have serial correlation, { and} dependent variable have cross section correlation, { when} $N\rightarrow\infty$, T is fixed, { and} $T/N\rightarrow0$ (short panel data), then
$$
\sqrt{N}[(\hat{\beta}_{L},\hat{\beta}_{F})-(\beta_{L},\beta_{F})]\xrightarrow{d}N(0,\Sigma_{1});
$$
(2) Explanatory variables have serial correlation, { and} dependent variable{ s do not have} cross section correlation, { when} $N, T\rightarrow\infty$, and $T/N\rightarrow C$, $C$ is constant (long panel data), $C\neq0$, then
$$
\sqrt{NT}[(\hat{\beta}_{L},\hat{\beta}_{F})-(\beta_{L},\beta_{F})]\xrightarrow{d}N(0,\Sigma_{2}).
$$
\end{theorem}

The proofs of { Theorem 2 can be seen in Appendix B}.

{ From Theorem 2 we can get the conclusion that} asymptotic normality of sample estimator for short panel ($T\ll N$) and long panel ($T$ and $N$ is close) { can be got}. The value{ s} of $\Sigma_{1}$ and $\Sigma_{2}$ { are correlated} closely with asymptotic variance $D_{1}$ and $D_{2}$ of $\sqrt{N}O_{N}$. Optimal Weighted matrix $A_{O}$ { is} generally substituted by a random given matrix to obtain $D_{1}$ and $D_{2}$, so $D_{1}$ and $D_{2}$ { are} mainly dependent on the variance of random error term. Furthermore, { we assume}
$$E(\epsilon_{it}\epsilon_{it+h})=0$$
and $Var(\epsilon_{it})=\sigma_{\epsilon}^{2}$, so variance of disturbance term influence asymptotic variance of estimator varied with the estimation method of given model. The choice of IV and weighted matrix also influence  asymptotic variance, If
$$E[\Delta \epsilon_{i}|Z_{i}]=0, \quad i=1,\cdots,N
$$
then the interactive effect of error term and IV aren't considered, { which} is more stronger than $E[\Delta\epsilon_{i}Z_{i}]=0$.

Obviously, choosing different IV $Z$ also influence asymptotic variance of $\sqrt{N}O_{N}$, furthermore $\Sigma_{1}$ and $\Sigma_{2}$, so different number of IV will get different estimation results. For GMM estimation, appropriate IV come{ s} from higher order lag terms and exogenous variables, so it is important to choose the order of lag terms. Meanwhile, if every estimator of parameter{ s} to be estimated have asymptotic normality, by Slutsky's lemma, the asymptotic properties of the sum of these estimator will be obtained.

\section{Simulation Study}

DMDFM { is concerned with} time series correlation and cross section correlation simultaneously. To reflect these two styles of correlation, simulation processes permit that common factors of error term have lag effects. Common factors being decomposed by explanatory variables have individual correlation as well as series correlation. Factor loadings mainly reflect individual correlation. High { dimensional case includes} a large number of explanatory variables, { and} we attempt to use minority common factors to extract information of explanatory variables to reduce dimension. So, { in the simulation, we should} consider not only correlation with explanatory variables, but also lag effects of explanatory variables in { these common factors}. Consider the following data generation process (DGP):
$$
y_{it}=\alpha_{1}+\beta_{l1}y_{it-1}+\beta_{f1}f_{1it}+\beta_{f2}f_{2it}+\gamma_{i1}g_{1t}+\gamma_{i2}g_{2t}+\epsilon_{it} \eqno(25)
$$
{ Compared} with model (5), DGP add some restriction conditions to reflect { existing issues in terms of five parts}: Interception; first order lag of dependent variable; common factors of covariates; common factors and factor loadings of error components; idiosyncratic errors. As mentioned above, we choose two common factors from { each factor group}.

{ Intercept terms are generated from} normal distribution:
$$\alpha_{1}\sim i.i.d. N(1,2)$$

To reflect series correlation, the error term of model (5) { are generated from} AR(1) processes:
$$\epsilon_{it}=\rho_{\epsilon}\epsilon_{i,t-1}+\eta_{it}$$
$$\rho_{\epsilon}\sim i.i.d.U(0.05,0.95)$$
$$\eta_{it}\sim i.i.d.N(0,1)$$
$$\epsilon_{i,0}=0$$
{ The errors in this part} represent idiosyncratic error generated { from} factors decomposition. From the factor decomposition { process} of equation (2), { we see that} the other part of error components reflect in common factors and factor loadings of error term. Assume common factors of error component retain lag factors, and { we express} them as AR(1) processes from different idiosyncratic errors. { The first order correlation coefficients are generated from} uniform distribution, two error components DGP can be written as:
$$g_{jt}=\rho_{jt}g_{j,t-1}+u_{jt} \qquad (j=1,2)$$
$$\rho_{jt}\sim i.i.d.U(0.05,0.95),\quad g_{j,0}=0$$
$$u_{jt}\sim i.i.d.N(0,1)$$
where factor loadings of error component { are} always generated { from} uniform distribution or normal distribution, { and} here we use uniform distribution.
$$\gamma_{ki1}\sim i.i.d.U(0.05,0.95)$$
$$\gamma_{ki2}\sim i.i.d.U(0.05,0.95)$$

Common factors extracted from explanatory variables should reflect correlation among individuals, periods and explanatory variables. Every common factor of different individuals retain main information of explanatory variables and idiosyncratic component of individuals. So the data generation process of each common factor consists of four parts: level term; error factors term; individual correlation component; error component, which can be generated { from}:
$$f_{kit}=a_{ki1}h_{k1t}+\gamma_{ki1}g_{1t}+\gamma_{ki2}g_{2t}+\zeta_{k1t}q_{i1}+\omega_{kit} \quad (k=1,2)$$
where level term { is composed} of an individual random coefficient multiplied by an AR(1) processes. First order auto-correlation coefficients and initial value of AR(1) processes have been given, { and} the others { are generated} from AR(1) processes. Two common factors DGP of explanatory variables are:
$$a_{ki1}\sim i.i.d.U(0.05,0.95)$$
$$h_{k1t}=\rho_{kh}h_{k,1,t-1}+\tau_{kh}$$
$$\rho_{1h}=0.4,h_{1,1,0}=0.2,\rho_{2h}=0.5,h_{2,1,0}=0.3$$
$$\tau_{kh}\sim i.i.d.N(0,1)$$
{ Random} error of common factors terms { are generated} from normal distribution:
$$\omega_{kit}\sim i.i.d.N(0,0.25)$$
{ Individual} correlation components { are generated} from spatial auto-regression SAR(1), which can be { written as}:
$$q_{i1}=\rho_{q}q_{i-1,1}+\nu_{q}$$
$$\rho_{q}\sim i.i.d.U(0.05,0.95), q_{0,1}=0.1$$
$$\nu_{q}\sim i.i.d.N(0,1)$$
{ The} coefficients of individual correlation components { are generated} from uniform distribution:
$$\zeta_{11t}\sim i.i.d.U(0.05,0.95)$$
$$\zeta_{21t}\sim i.i.d.U(0.05,0.95)$$
{ The} common factors of explanatory variables retain the common factors of error components to express extracted information, whose coefficients { are generated} from uniform distribution:
$$\gamma_{i1},\gamma_{i2}\sim i.i.d.U(0.05,0.95)$$

Based on { the} above thoughts, { we} should give an initial value of the explanatory variables $y_{it}$:  $y_{i0}=0$, and $\beta_{l1}=0.6, \beta_{f1}=0.8, \beta_{f2}=1$. To ensure the { consistency} of the data generation process, we discarded the first 15 simulation value{ s}. Every experiment was replicated 2000 times for the (N,T)=(20,5), (50,5), (50,10), (100,5), (100,10), (100,20), (200,5), (200,10), (200,20), (200,50) respectively. The estimation results of parameters $\beta_{l1}, \beta_{f1}$ and $\beta_{f2}$ { are derived} from 2,000 times { replication}, whose mean bias and root mean square error (RMSE) are calculated hereafter. The simulations results are summarized in Tables 1.

\begin{center}
{{\bf Table 1}. Bias and RMSE of simulation results}
\end{center}
\begin{center}
{
\setlength{\tabcolsep}{1.5mm}


             \vskip 0.5cm

\begin{tabular}{lccccccccccccccc}\hline\hline

\multirow{2}{*}{(N,T)}& & \multicolumn{3}{c}{Bias}  & &
\multicolumn{3}{c}{RMSE} \\

\cline{3-5} \cline{7-9}

 && $\beta_{l1}$&$\beta_{f1}$&$\beta_{f2}$&& $\beta_{l1}$&$\beta_{f1}$&$\beta_{f2}$\\
\hline
   (20,5)&&     -0.0981&	-0.0333&	-0.0319&&	0.01238&	0.01313&	0.01301\\
   (50,5)&&	    0.00131&	0.02163&	0.00161&&	0.00962&	0.01013&	0.01036\\
   (50,10)&&    0.03613&	0.00423&	0.02274&&	0.00377&	0.00655&	0.00647\\
   (100,5)&&	0.01293&	0.02180&	0.01876&&   0.00942&	0.00909&	0.00914\\
   (100,10)&&	0.04707&	0.00758&	0.01898&&	0.00361&	0.00607&	0.00581\\
   (100,20)&&	0.07718&	0.00616&	0.01829&&	0.00189&	0.00459&	0.00417\\
   (200,5)&&	0.03012&	0.02184&	0.02596&&   0.00994&	0.00808&	0.00888\\
   (200,10)&&	0.05712&	0.00829&	0.01733&&	0.00364&	0.00591&	0.00565\\
   (200,20)&&	0.08152&	0.00812&	0.02530&&	0.00187&	0.00432&	0.00425\\
   (200,50)&&	0.08032&	0.01192&	0.02292&&   0.00187&	0.00448&	0.00420\\

\hline\hline
\end{tabular}}
\end{center}

As can be seen from Table 1, { when the values of $N$ and $T$ are given, the} first order lag term of dependent variable in DMDFM { has} smaller bias and RMSE as well as coefficient estimation value of explanatory variables' common factors. It indicates that GMM estimation can obtain consistent and efficient parameter estimator. Furthermore, { considering} the size of relative bias, we can see that the range of dependent variable and explanatory variables are in  (-20,20). The results of table 1 is relative smaller than initial values, so the estimators are consistent correspondence with population parameter. These satisfy the large sample properties of DMDFM and GMM estimation { mentioned} previously.

The results of simulation demonstrate that bias and RMSE of regression coefficient{ s} do not obviously vary with the number of individuals increment. The number of individuals increase from 20 to 200 and periods increase from 5 to 50, but bias do not increase with individual size, which indicates { that} the results of estimation have good properties { in} finite sample. For the short panel, periods { are} shorter than the number of individual{ s}, bias and RMSE s do not vary obviously. When periods become longer, according to the results of Monte Carlo simulation, the bias of estimator become smaller. In { the} above simulation, the number of individuals { is} at least 4 times more than periods length, which { reveals} the high dimensional feature. Bias of estimator become{ s} smaller with { the} increasing the periods length. The results of estimation have higher uniform convergence speed.

Root Mean Square Error (RMSE) include{ s} the information of sample bias and variance. The results of Table 1 demonstrate that bias and RMSE are smaller. { It shows} that the variance is also smaller due to { that} MSE is the sum of variance and square of bias. The smaller variance of estimation results indicate that this estimation method { can} not only obtain consistency estimator but also obtain efficient variance. This { verified} consistency and efficiency once again.

DMDFM can reduce the dimension of indicators and reflect the internal structure of panel data reasonably{ . Furthermore,} the model estimation results can be used for predicting dependent variable. In order to test the prediction effect of DMDFM, we still { use} the DGP as { before} to generate a group training { sets} and testing { sets}. To enhance the observability of the graphics, we predict 20 periods values of dependent variable step by step. At first { we} generate every periods value of explanatory variables and one period lag value of dependent variable, then predict dependent variable forward one period by two step estimation method through model (5)-(7) to compare the predicted values and true values. Figure 1 { shows the average of} predicted value{ s} of 100 individual{ s} compared with true value{ s}. Figure 2 is 6 individuals which extracted randomly from 100 individual predicted values compared with true values.

As can be seen from Figure 1 and Figure 2, predicted  values of all individual average and every individual have good prediction effect via GMM estimation. One step predicted value have goodness fitting of trend as well as points. The constructed model and its estimation method reflect the data generation processes well, and prediction effect is better. Furthermore, if we consider Mean Absolute Percentage Error (MAPE), the similar conclusion should be obtained.

\begin{figure}[htbp]
\includegraphics[bb=25 250 0 600,scale=.75]{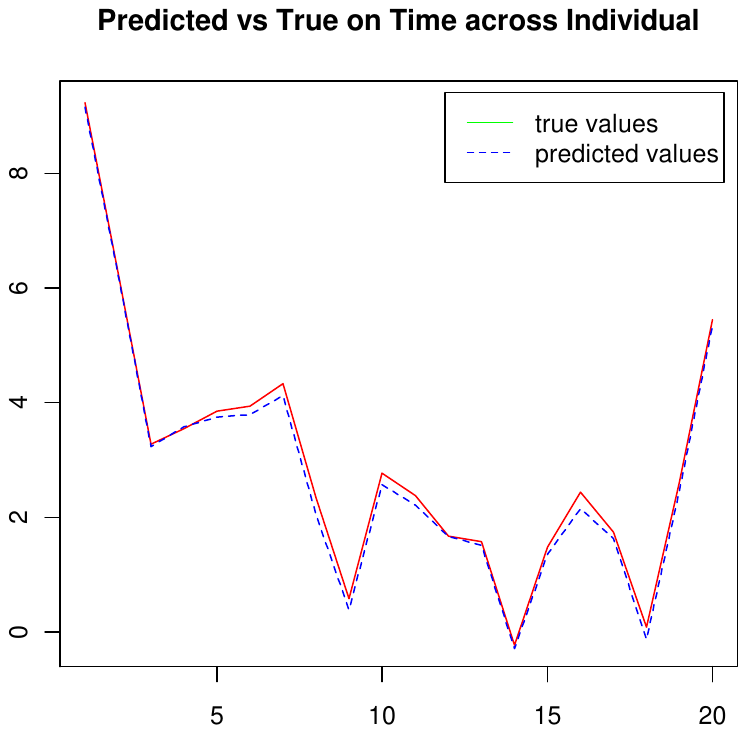}
\caption{Average predicted and true value on 20 periods across 100 individuals}
\end{figure}

\begin{figure}[htbp]
\includegraphics[bb=25 250 0 600,scale=.75]{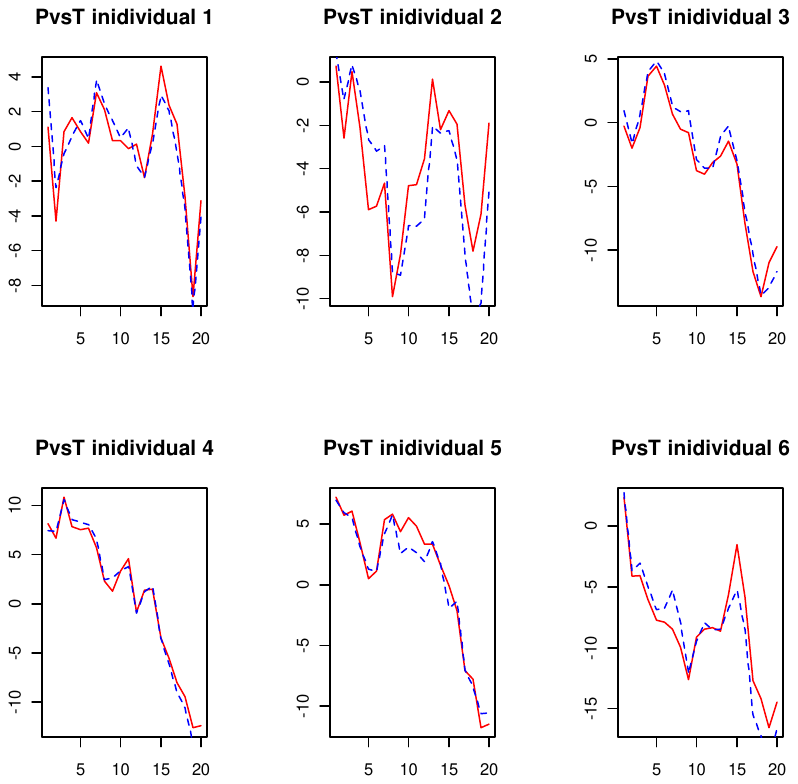}
\caption{Predicted and true value on 20 periods across 6 individuals}
\end{figure}

\section{ Conclusion}

In this article, we propose a panel data double factors model which { include} both explanatory variables and error component factor decomposition. The Mixed Double Factor Model derive{ s} from the factor decomposition method, { and the aim of twice decomposition analysis} is different. Contrast to the general dynamic factor model, the dynamic of DMDFM refer to the lag terms of dependent variable. { Theoretically}, if panel data have first order auto correlation of time series and heterogeneity components of individual or periods (fixed effect of random effect), the lag term $Y_{i,t-1}$ of dependent variable $Y_{it}$ are determined by the expectation of two parts information: the lag information sets $I_{t-1}$ of explanatory variables $X_{it}$ and the remainder information given by $X_{it}$, i.e. $E(Y_{it})=E(Y_{it}|I_{t-1},X_{it})=E(Y_{it}|Y_{it-1},X_{it})$. The dynamic panel data model{ s} constructed by lag terms of explanatory variables and common factors are different, however the results are excellent. Dynamic Mixed Double Factor Model { is composed of} four main parts: lags of responsor; common factor of regressors; factor error component and idiosyncratic error.

RHS of dynamic panel data model include{ s} the lag terms of dependent variable, so independent assumption of error term and dependent variable aren't satisfied. { We cannot get the consistent and efficient estimators using }OLS or MLE of dynamic factor model, { so} generalized moment method (GMM) is a better alternative options. In this article, we propose an iteration GMM to estimate DMDFM. At first, we obtain { the} error component of the model through GMM estimation, furthermore decompose factor with the given error component. The factors decomposition results of estimated error component can be regarded as { intercept term} of new model which can be estimated by GMM to obtain parameter estimation value once again. The proof of the { Theorems} and simulation results show that the two-step GMM estimation { is} able to get consistent estimators of the dynamic Mixed Double Factor Model. The estimation results of DMDFM have better explanatory power and prediction effects.

DMDFM reduce{ s} the dimension of large number of indicators. In which, a large number of explanatory variables { are} represented by few common factors, { which} extends the application scope of the model. However, every explanatory variable has its own implication in empirical analysis, { and} we should consider how to provide reasonable explanation of explanatory variables in { the following} step. The research scope of this article only aims at dimensional reduction, { while} variable selection for the explanatory ability of the indicators is not considered, which restrict the application effect of the model. Panel data usually { has} serial correlation and cross-section correlation, { and} there perhaps exist{ s} other structural features. These structural features related to individual{ s} are obvious in the spacial panel data, i.e. structural change, heteroscedasticity and variance magnitude, and so on, however DMDFM can not solve these problem thoroughly. We will study how to improve DMDFM to reflect the structural features of panel data in the future.

The estimators of DMDFM mainly focus on expectation in this article, however variance of DMDFM also should be taken into account as well as multivariate time series heteroscedasticity model. { Other} issues of DMDFM include: consistent asymptotic variance estimation; asymptotic efficiency of estimators; testing of estimators obtained by GMM estimation, etc. In addition to { theoretical} analysis of model construction and estimation, empirical research also should be considered. Because high dimensional panel data appear{ s} both in macroeconomic and microeconomic fields, empirical research combine{ d} with application background must be discussed in future.

\section*{Appendix:  Proof of Theoretical results}

\noindent {\bf A. Proof of Theorem 1.}

Denote $b(z,\beta)=Z_{i}\Delta\epsilon_{i}$, where $\beta=(\beta_{L}^{'},\beta_{F}^{'})_{'}$. From equation (16), we have $E[b(z,\beta)]=0$. { we} calculate partial derivative for each parameter to be estimated, $\partial b(z,\beta)/\partial\beta$, then let
$$Db(\beta_{L},\beta_{F})=(\partial b(b(z,\beta)/\partial\beta_{L}^{'},\partial b(z,\beta)/\partial\beta_{F}^{'})^{'}$$
because the uniform consistency of random disturbance term, using Taylor series expansion around $\beta_{L}$ and $\beta_{F}$:
$$b(z,\hat{\beta})=b(z,\beta)+Db(\beta_{L}^{*},\beta_{F}^{*})(b(z,\hat{\beta})-b(z,\beta))+o(b(z,\beta)) \quad \eqno(A.1)$$
where $\hat{\beta}=(\hat{\beta}_{L}^{'},\hat{\beta}_{F}^{'})^{'}$, $\beta_{L}^{*}$, $\beta_{F}^{*}$ { are} between $\beta_{L}$, $\hat{\beta}_{L}$, and $\beta_{F}$, $\hat{\beta}_{F}$ respectively, multiplied { by} weighting matrix $A$ simultaneously:
$$ Ab(z,\hat{\beta})=A b(z,\beta)+ADb(\beta_{L}^{*},\beta_{F}^{*})(b(z,\hat{\beta})-b(z,\beta))+o(b(z,\beta)) \quad \eqno(A.2)$$
{ Given the following three items}:

(i) From assumption{ s} as before, given optimal weighting matrix $A_{O}$, { we} can obtain unique optimal estimator of $\beta$. $\beta$ is continuous vector definite{ d} on Euclid space $R^{n}$, { and} space $\Theta$ constituted by $\beta$ is a subset of $R^{n}$, and is closed and bounded.

(ii) For $b(z,\beta)=Z_{i}\Delta\epsilon_{i}$, $\forall \epsilon>0$, from (A.1)
$$E(b(z,\hat{\beta}))=b(z,\beta)$$
so,
$$|b(z,\hat{\beta})-b(z,\beta)|\xrightarrow{p}0 \quad \eqno(A.3)$$
for given matrix $A$, denote
$$\hat{S}_{N}(\beta)=b(z,\hat{\beta})^{'}\hat{A}b(z,\hat{\beta})$$
and
$$S_{0}(\beta)=b(z,\beta)^{'}Ab(z,\beta)$$
from (A.3), $S_{0}(\beta)$ is continuous.

(iii) Next, prove $S_{0}(\beta)$ convergence with probability 1.
$$|\hat{S}_{N}(\beta)-S_{0}(\beta)|=|b(z,\hat{\beta})^{'}\hat{A}b(z,\hat{\beta})-b(z,\beta)^{'}Ab(z,\beta)|$$
$$=|(b(z,\hat{\beta})-b(z,\beta))^{'}\hat{A}(b(z,\hat{\beta})-b(z,\beta))$$
$$+b(z,\beta)^{'}\hat{A}(b(z,\hat{\beta})-b(z,\beta))+b(z,\beta)^{'}\hat{A}b(z,\beta)-b(z,\beta)-b(z,\beta)^{'}Ab(z,\beta)|$$
$$=|(b(z,\hat{\beta})-b(z,\beta))^{'}\hat{A}(b(z,\hat{\beta})-b(z,\beta))$$
$$+b(z,\beta)^{'}\hat{A}(b(z,\hat{\beta})-b(z,\beta))^{'}\hat{A}(b(z,\beta)+b(z,\beta)^{'}(\hat{A}-A)b(z,\beta)|$$
$$=|(b(z,\hat{\beta})-b(z,\beta))^{'}\hat{A}(b(z,\hat{\beta})-b(z,\beta))$$
$$+b(z,\beta)^{'}(\hat{A}+\hat{A}^{'})(b(z,\hat{\beta})-b(z,\beta))+b(z,\beta)^{'}(\hat{A}-A)b(z,\beta)|$$
Using triangle inequalities
$$\leq|(b(z,\hat{\beta})-b(z,\beta))^{'}\hat{A}(b(z,\hat{\beta})-b(z,\beta))|$$
$$+|b(z,\beta)^{'}(\hat{A}+\hat{A}^{'})(b(z,\hat{\beta})-b(z,\beta))|+|b(z,\beta)^{'}(\hat{A}-A)b(z,\beta)|$$
Using Cauchy-Schwartz inequalities
$$\leq\|b(z,\hat{\beta})-b(z,\beta)\|^{2}\|\hat{A}\|$$
$$+2\|b(z,\beta)\|\|b(z,\hat{\beta})-b(z,\beta)\|\|\hat{A}\|+\|b(z,\beta)\|^{2}\|\hat{A}-A\|$$
because
$$b(z,\hat{\beta})-b(z,\beta)\xrightarrow{p}0$$
$$\hat{A}-A\xrightarrow{p}0$$
we have
$$|\hat{S}_{N}(\beta)-S_{0}(\beta)|\xrightarrow{p}0$$

By Newey and McFadden (1994), { following} uniform convergence theorem, the conclusion is obtained.

\noindent {\bf B. Proof of Theorem 2.}

(1) Because
$$\partial R_{1}(\beta_{L},\beta_{F})/\partial \beta=\partial(b(z,\beta)^{'}Ab(z,\beta))/\partial \beta$$
$$=\partial(b(z,\beta)^{'}/\partial\beta Ab(z,\beta))+\partial(b(z,\beta)^{'}/\partial\beta Ab(z,\beta))$$
$$=2\partial(b(z,\beta)^{'}/\partial\beta Ab(z,\beta))$$
where $\beta=(\beta_{L},\beta_{F})^{'}$ for { notation} simplicity. { Following }this notation, { in order to estimate GMM}, we solve first order condition, so { we} obtain { that}
$$R_{1}(\hat{\beta})^{'}Ab(z,\hat{\beta})=0 \eqno(B.1)$$
from (A.1), for optimal matrix $A_{O}$, we have
$$R_{1}(\beta)^{'}A_{O}b(z,\hat{\beta})=R_{1}(\beta)^{'}A_{O}\sqrt{N}b(z,\hat{\beta})+o(b(z,\beta)) \eqno(B.2)$$
using Taylor series expansion around $\beta$
$$R_{1}(\beta)^{'}A_{O}b(z,\hat{\beta})=R_{1}(\beta)^{'}A_{O}(\sqrt{N}b(z,\beta)+R_{1}(\beta)\sqrt{N}(\hat{\beta}-\beta))+o(b(z,\beta))$$
from (B.1),we have
$$R_{1}(\beta)^{'}A_{O}R_{1}(\beta)\sqrt{N}(\hat{\beta}-\beta)=-R_{1}(\beta)^{'}A_{O}\sqrt{N}b(z,\beta)+o(b(z,\beta))$$
so
$$\sqrt{N}(\hat{\beta}-\beta)=-(R_{1}(\beta)^{'}A_{O}R_{1}(\beta))^{-1}R_{1}(\beta)^{'}A_{O}\sqrt{N}b(z,\beta)+o(b(z,\beta))$$
by equation (23) as previous, we have
$$\sqrt{N}b(z,\beta)\xrightarrow{d}N(0,D_{1})$$
and
$$(R_{1}(\beta)^{'}A_{O}R_{1}(\beta))^{-1}R_{1}(\beta)^{'}A_{O}$$
is a determined matrix, so
$$\sqrt{N}(\hat{\beta}-\beta)\xrightarrow{d}N(0,\Sigma_{1})$$
i.e.
$$\sqrt{N}((\hat{\beta}_{L},\hat{\beta}_{F})-(\beta_{L},\beta_{F}))\xrightarrow{d}N(0,\Sigma_{1})$$

\begin{flushright}
Q. E. D.
\end{flushright}

(2) The proof is similar with (1), with the same argument, we can prove it.

\vspace{3ex}
\section*{Acknowledgements}
\noindent Zhang's research was partially supported by the
Fundamental Research Funds for the Central Universities,  the
Research Funds of Renmin University of China (10XNL007). Fang's
research was partially supported by the NSFC (71271210,714711730).

\end{document}